\documentclass[conference]{IEEEtran}
\IEEEoverridecommandlockouts
\usepackage{cite}
\usepackage{amsmath,amssymb,amsfonts}
\usepackage{algorithmic}
\usepackage{graphicx}
\usepackage{textcomp}
\usepackage{xcolor}
\usepackage{subfig}
\def\BibTeX{{\rm B\kern-.05em{\sc i\kern-.025em b}\kern-.08em
    T\kern-.1667em\lower.7ex\hbox{E}\kern-.125emX}}
\begin{document}

\title{Improving Real-time Communication for Educational Metaverse by Alternative WebRTC SFU and Delegating Transmission of Avatar Transform}

\author{\IEEEauthorblockN{Yong-Hao Hu}
\IEEEauthorblockA{\textit{Virtual Reality Educational Research Center} \\
\textit{The University of Tokyo}\\
Tokyo, Japan \\
yh-haoareyou@cyber.t.u-tokyo.ac.jp}
\and
\IEEEauthorblockN{Kenichiro Ito}
\IEEEauthorblockA{\textit{Virtual Reality Educational Research Center} \\
\textit{The University of Tokyo}\\
Tokyo, Japan \\
ito@vr.u-tokyo.ac.jp}
\and
\IEEEauthorblockN{Ayumi Igarashi}
\IEEEauthorblockA{\textit{Graduate School of Medicine} \\
\textit{The University of Tokyo}\\
Tokyo, Japan \\
aigarashi-tky@g.ecc.u-tokyo.ac.jp}
}

\IEEEaftertitletext{\vspace{-2\baselineskip}}

\maketitle

\begin{abstract}
Maintaining real-time communication quality in metaverse has always been a challenge, especially when the number of participants increase.
We introduce a proprietary WebRTC SFU service to an open-source web-based VR platform, to realize a more stable and reliable platform suitable for educational communication of audio, video, and avatar transform.
We developed the web-based VR platform and conducted a preliminary validation on the implementation for proof of concept, and high performance in both server and client sides are confirmed, which may indicates better user experience in communication and imply a solution to realize educational metaverse.

\end{abstract}


\section{Introduction}

As the rising use of metaverse, usually referred to 3D virtual space accessible from electronic devices including computers or VR devices, in various situations, its application in education has also been conducted and discussed\cite{b1}\cite{b2}.
Similar to how Moodle\footnote{https://moodle.org/} serves as an representative Learning Management System (LMS), we consider it essential to build an inclusive platform in metaverse for education.
We chose Mozilla Hubs\footnote{https://hubs.mozilla.com/} as the base for such platform due to its open-source-driven, highly accessible browser-based nature, complete architecture and well-proven track record.

Mozilla Hubs currently has a limit of 24 participants per room\footnote{https://support.mozilla.org/en-US/kb/room-capacity-hubs}, and raising the limit decreases the communication quality, since real-time communication within 3D virtual spaces includes not only media (audio and video) but also spatial data such as avatar transform, which results in much heavier data transmission especially with high number of participants.
Creating multiple rooms and redirecting participants is an alternative to accommodate more participants, however, this approach may not be suitable for an inclusive online classroom where all participants may attend in the same room.

The goal of this study includes achieving comparable or higher communication quality through audio, video, and spatial data in the same room without being impacted by the number of clients per room, for the purpose of developing an inclusive educational metaverse.

\section{Development of Inclusive Educational Metaverse Platform}

\vspace{-2pt}
\subsection{Requirements}

We considered that the basis being open-source, browser-based, and capable of maintaining communication quality are required for an education platform to be sustainable and inclusive.
An open-source software, depending little on proprietary ones, are more flexible to customize and robust against discontinuation of proprietary software, while proprietary software could still be utilized as tools with benefits to reduce the cost of operation and maintenance, such as Google Workspace or Microsoft 365.
A browser-based service, available for various devices with browsers installed, enhances the accessibility for users within diverse situations.
In contrast, services being not browser-based and depending on any specific operating system require concurrent maintenance to comply with the OS update life cycle, which many applications failed to keep up with.

Stable and reliable communication is indispensable for education, and low communication quality owing to the number of students is undesirable.
Current number of students per classroom in Japan is up to around 35 or 40, and we expect an online classroom to support at least the same capacity while keeping its communication quality.
Therefore, educational metaverse should not merely be open-source and browser-based, which are part of the reasons we adopt Mozilla Hubs, but also maintain communication quality with little impact from the number of participants, which is what we attempts to improve by utilizing proprietary software as an alternative.

\vspace{-2pt}
\subsection{Method}

Mozilla Hubs' original architecture\footnote{https://hubs.mozilla.com/docs/system-overview.html} transmit spatial data through WebSocket on a mesh network, and transmit media using their own WebRTC Selective Forwarding Unit (SFU) named Dialog, which was formerly based on Janus\footnote{https://janus.conf.meetecho.com/} and currently Mediasoup\footnote{https://mediasoup.org/}, both open-source WebRTC SFU libraries.

We propose introducing an alternative WebRTC SFU solution besides Mediasoup for media transmission, and delegate the transmission of avatar transform to the alternative.

\subsection{Implementing WebRTC SFU Alternative}
Sora, a WebRTC SFU provided by Shiguredou Inc. in Japan\footnote{https://sora.shiguredo.jp/} was chosen for implementation due to its featured higher client capacity (1:1000 broadcasting per room), conciser implementation, swift response to new browser updates, and more stable signaling through a hybrid of WebSocket and DataChannel.

\begin{figure}[t]
\vspace{-4pt}
\centerline{\includegraphics[width=7cm]{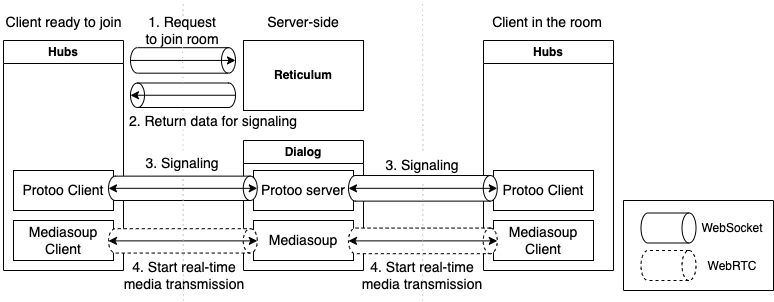}}
\vspace{-4pt}
\caption{Mozilla Hubs architecture with WebRTC SFU}
\label{dialog}
\vspace{-12pt}
\end{figure}

In current architecture of Mozilla Hubs, a client joining a room uses data retrieved from Reticulum, the web server of Mozilla Hubs, to conduct signaling through Protoo\footnote{https://protoo.versatica.com/} to create WebRTC connections with other clients, and media starts being transmitted through Mediasoup, as shown in Fig.~\ref{dialog}.

\begin{figure}[t]
\centerline{\includegraphics[width=7cm]{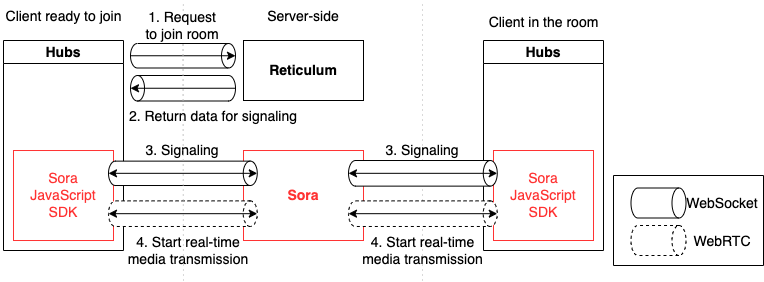}}
\vspace{-4pt}
\caption{Proposed architecture with proprietary WebRTC SFU}
\label{sora}
\vspace{-16pt}
\end{figure}

Our implementation is shown in Fig.~\ref{sora}.
Components in red are the differences from the default Mozilla Hubs architecture.
Sora is capable of handling signaling, which Mediasoup is not\footnote{https://mediasoup.org/documentation/v3/communication-between-client-and-server/}, therefore, Sora handles both signaling and data transmission for the proposed implementation.
In our implementation, we host Reticulum, Dialog and Hubs frontend on the same server, and when a room manager chooses to use Sora, Dialog is paused and Sora's cloud service starts serve as the WebRTC SFU.
Hence, the implementation provides a proprietary software solution to easily solve existing problem when needed, but also avoids the software to have dependency on it.

\subsection{Delegate Avatar Transform Transmission to WebRTC SFU}
Current architecture of Mozilla Hubs synchronizes avatar transform through Reticulum, the web server, using the reliable WebSocket protocol.
However, real-time synchronization between voice and avatar poses may be regarded more important for communication than simply prioritizing reliability.

WebRTC Datachannel was implemented on top of UDP while remaining reliable as TCP, suitable for the real-time transmission of avatar transform data.
The transmission of avatar transform was delegated to Sora's Datachannel, including position and rotation of body, head, and hands.

\section{Preliminary Validation: Proof of Concept}
A preliminary validation was conducted on implementation with Sora, compared with the original Dialog implementation.
For each condition, 12 devices typically used at educational scenery was connected: 1 Apple Mac, 1 Laptop Windows 10, 2 iPad, 1 Chromebook, 1 Microsoft Surface, 3 iPhone, 1 Android smartphone, 1 Meta Quest 2, and 1 Pico 4.
Server data and client data was obtained for 5 minutes, no severe delay in media and avatar transform was observed for both conditions.

\begin{figure}[h]
\centering
\vspace{-20pt}
\subfloat
{{\includegraphics[width=3.5cm]{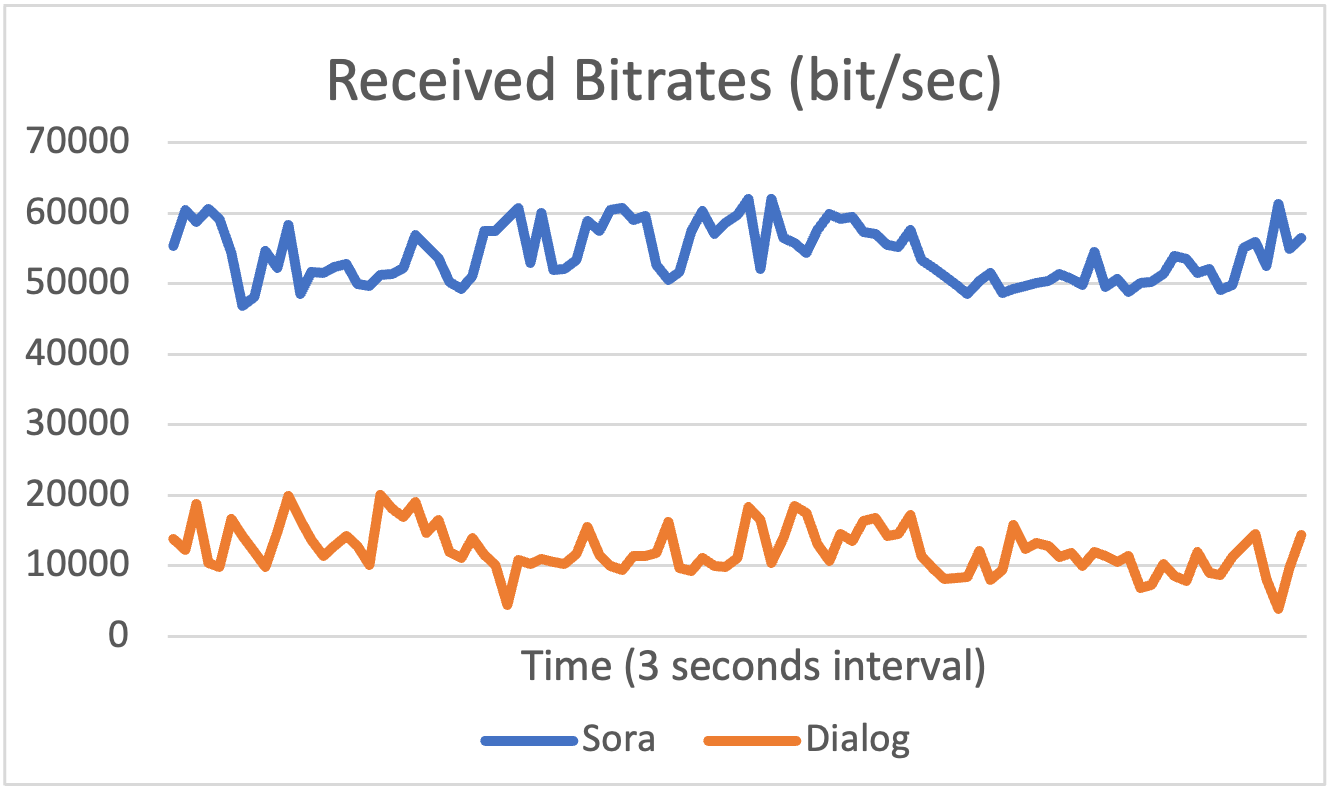} }}%
\subfloat
{{\includegraphics[width=3.5cm]{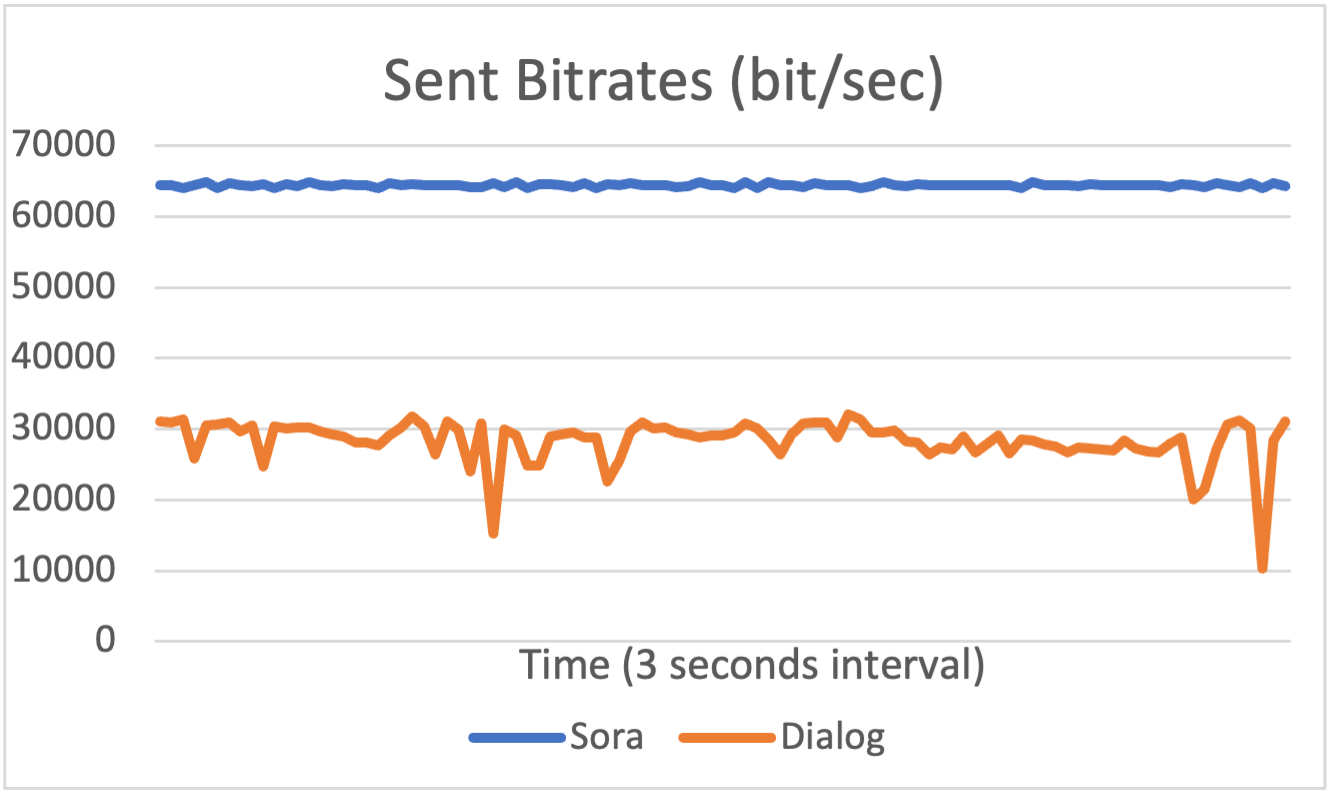} }}%
\vspace{-4pt}
\caption{WebRTC bitrates recorded every 3 seconds}
\label{bitrate-by-time}
\vspace{-6pt}
\end{figure}

Server's average load was collected every minute (Table~\ref{load}).
The results indicate that the proposed implementation using Sora relieves server load than the original implementation.
Client data of transmitted bytes between clients with WebRTC getStats API was collected on the Apple Mac device.
Bitrates were calculated and plotted in Fig.~\ref{bitrate-by-time}, with the average listed in Table~\ref{bitrate}.
The results indicate higher sent/received and stabler sent bitrates for implementation with Sora.

\begin{table}[h]
\caption{Server average load (number of processes waiting for CPU time)}
\vspace{-12pt}
\begin{center}
\begin{tabular}{c c c}
\hline
 & \textbf{\textit{Sora}}& \textbf{\textit{Dialog}} \\
\hline
Average load over 1st minute & 0.17 & 0.15 \\
Average load over 2nd minute & 0.06 & 0.13 \\
Average load over 3rd minute & 0.02 & 0.09 \\
Average load over 4th minute & 0.01 & 0.09 \\
Average load over 5th minute & 0.03 & 0.08 \\
\hline
\hline
Average load over 5 minutes & 0.058 & 0.108 \\
SD of average load over each minute & 0.0584 & 0.0271 \\
\hline
\end{tabular}
\label{load}
\end{center}
\vspace{-12pt}
\end{table}

\begin{table}[h]
\caption{Average WebRTC bitrates in 5 minutes}
\vspace{-12pt}
\begin{center}
\begin{tabular}{c c c c c}
\hline
 & \multicolumn{2}{c}{\textbf{Received (bit/sec)}} & \multicolumn{2}{c}{\textbf{Sent (bit/sec)}} \\
 & \textbf{\textit{Sora}}& \textbf{\textit{Dialog}} & \textbf{\textit{Sora}}& \textbf{\textit{Dialog}} \\
\hline
Average & 54204.932 & 12234.112 & 64399.084 & 28333.127 \\
SD & 3970.0196 & 3262.1003 & 225.6512 & 3115.1473 \\
\hline
\end{tabular}
\label{bitrate}
\end{center}
\vspace{-12pt}
\end{table}

\section{Conclusion}
We introduced a proprietary WebRTC SFU service to Mozilla Hubs to improve media and avatar transform transmission.
Preliminary validation results showed less server load and higher bitrates, which implies better user experience in communication and feasibility of a metaverse for education.

\section*{Acknowledgment}
This work was partially supported by the following grants: JST Grant Number JPMJPF2202, JSPS KAKENHI Grant Number 22K19683.


\begin{thebibliography}{00}
\bibitem{b1} R. Alfaisal, H. Hashim, and U.H. Azizan, ``Metaverse system adoption in education: a systematic literature review,'' J. Comput. Educ. December 2022.
\bibitem{b2} H. Lin, S. Wan, W. Gan, J. Chen and H. Chao, ``Metaverse in Education: Vision, Opportunities, and Challenges,'' IEEE Int. Conf. Big Data. Japan, pp. 2857--2866, 2022
\end{thebibliography}
\end{document}